\documentclass[a4paper,12pt]{article}
\usepackage[cp1250]{inputenc}
\usepackage{epsfig}
\usepackage{amssymb}
\usepackage[english]{babel}
\usepackage{amsmath}
\usepackage{setspace}
\usepackage{color}
 0
\title{On quantum corrections to dislocations mass}
\author{Grzegorz Kwiatkowski and Sergey Leble,\\ leble@mifgate.pg.gda.pl,  gkwiatkowski@mifgate.pg.gda.pl\\
\\
  Gda\'nsk University of Technology, \\
ul. G. Narutowicza 11/12, 80-952 Gda\'nsk, Poland\\}

\definecolor{blue}{rgb}{0,0,1}
\definecolor{black}{rgb}{0,0,0}
\begin{document}
 \maketitle
\renewcommand{\abstractname}{\small }

 \begin{abstract}
Quasiclassical quantization of crystal dislocations field is
considered in terms of functional integral.  The
generalized zeta-function is used to evaluate the functional
integral and  quantum corrections  to mass in quasiclassical
approximation.
The quantum corrections to few classical solutions of one-dimensional
Sin-Gordon model  are evaluated with account of rest $n-1$ dimensions.
The results are applied to appropriate crystal dislocation models.
\end{abstract}

\section{Introduction}
	We consider  Frenkel-Kontorova models for crystal structure dislocations
\cite{Br,Braun} and its Sin-Gordon (SG) equation counterpart. Being a one-dimensional field theory, which continuous limit is based on nonlinear Klein-Fock-Gordon (KFG) equations, it should be embedded into two- or three-dimensional picture with effective account of rest variables as in the model itself as in quantization procedure.

	The history of kink/soliton field quantization, started from \cite{KF,Das,Raj} is rather impressive now \cite{BJV,Mach}. One of advanced method, that allows to obtain relevant results in compact form is reviewed in \cite{Raj}, the  semi-classical quantization is a well known method of approximating energies of nonlinear equation's states.

	Beginning from Maslov paper \cite{Mas} the functional integral method becomes practical tool for evaluation of quasiclassical corrections to the action. At the meantime in \cite{KF} and in \cite{Das} such expressions were studied for SG and $\phi^4$  models: approximate quantum corrections were evaluated. The problem of embedding such model into real multidimensional theory and the problem of regularization is still under examination \cite{Kon} \cite{kwant}, \cite{ZL}. Generalization for the supersymmetric kink is given in   \cite{Bor}.

    Aim of this work is to apply the semi-classical quantisation method to calculate energy correction for Sin-Gordon kinks in real crystals.  A
general algebraic method of quantum corrections evaluation based on zeta-function \cite{RS} is used and  the Green function for heat equation with soliton potential is constructed. The mentioned problems of the theory dimensions and regularization are traced.

     In the first section connection between Frenkel-Kontorova model for a real solid and Sin-Gordon equation will be given and the kink solution of Sin-Gordon equation is reproduced with the realistic parametrization scheme. In the next section a space-time approach to the continual integral evaluation \cite{Feyn} and its semi-classical approximation of energy for static field equation's solutions will be described as well as the zeta-function regularisation method. Following section will be devoted to calculating corrections to energy of Sin-Gordon kink with expressions via constants of some solids. In last two sections these corrections  are linked to Frenkel-Kontorova model of dislocations and crowdions.

\section{Frenkel-Kontorova model and Sin-Gordon equation}\label{FKSG}

\indent Frenkel-Kontorova model describes one-dimensional chain of atoms interacting with each other and with a sinusoidal potential. Motion equation can be written as \cite{Braun}
\begin{equation}\label{FK}
    M\frac{\partial^2 x_i}{\partial t^2}=G\left(x_{i+1}-2x_i+x_{i-1}\right)-\frac{d V}{dx}(x_i),
\end{equation}
where $V$ (the external potential) usually takes the form
\begin{equation}
    V(x)=\frac{\varepsilon}{2}\left(1-\cos\left(\frac{2\pi x}{a}\right)\right),
\end{equation}
where $M$ is the mass of atoms of which the particular crystal lattice is built, $a$ is the lattice constant, $G$ is proportional to the bulk modulus $K$ of a given material ($G=Ka$) and $\varepsilon$ is proportional to shear modulus $M_s$ ($\varepsilon=\frac{2a^3 M_s}{\pi^2}$ - only interaction with nearest neighbors is taken into account).
It is also important, to note, that $x_i$ can be viewed as actual position of particular atoms or deviation from equilibrium position (if equilibrium distance between atoms is equal to the external potential's period). The latter interpretation will be useful for obtaining continuous approximation (Sin-Gordon equation). Let $u$ be a function of real variable $x$ and let it solves the equation (\ref{FK}) on condition:
\begin{equation}
    u(i a,t)=x_i, \qquad \forall i\in \mathbb{Z}.
\end{equation}
By expanding all instances of function $u$ in a Taylor series around $ia$ and omitting higher derivatives of $u$, one obtains Sin-Gordon equation
\begin{equation}
    \frac{M}{a}\frac{\partial^2 u}{\partial t^2}-a G\frac{\partial^2 u}{\partial x^2}+\frac{\varepsilon \pi}{a^2}\sin\left(\frac{2\pi u}{a}\right)=0.
\end{equation}
For the purpose of this work, we will rewrite the equation in unitless variables, with $T$ as a parameter connected with action $S$ (specified in the next paragraph):
\begin{equation}
    x=ax',
\end{equation}
\begin{equation}\label{zamiana}
    u=a\phi ,
\end{equation}
\begin{equation}
    t=Tt'.
\end{equation}
That yields
\begin{equation}\label{not}
\frac{M}{T^2}\frac{\partial^2 \phi}{\partial t'^2}-G\frac{\partial^2 \phi}{\partial x'^2}+\frac{\varepsilon \pi}{a^2}\sin\left(2\pi \phi\right)=0.
\end{equation}
Note, that the change of sign in front of cosine in the potential only moves the solution by a constant, thus doesn't have any notable effect. Well known static elliptic solution \cite{ZL} of Sin-Gordon equation can be expressed as
\begin{equation}\label{phi}
    \phi=\frac{1}{\pi}\arcsin\left(k\ sn\left(m x';k\right)\right)+\frac{1}{2}.
\end{equation}
with $m=\sqrt{\frac{2\varepsilon}{a^2 G}}\pi$, $k\in [0,1]$. Kink case is obtained for $k=1$. Classical kink energy in notation of (\ref{not}):
\begin{equation}
    E_c=\frac{1}{\pi}\sqrt{8\varepsilon a^2 G}.
\end{equation}

\section{Method examination}
Let us consider a one-dimensional nonlinear Klein-Gordon-Fock (KFG) equation with $t'$ as time, $x'$ as spatial coordinate and  a twice-differentiable potential $\frac{\varepsilon}{2a}\left(1-cos\left(2\pi \varphi(x')\right)\right)$
\begin{equation}
    \frac{aM}{T^2}\frac{\partial^2 \varphi}{\partial t'^2}=aG\frac{\partial^2 \varphi}{\partial x'^2}-\frac{\varepsilon\pi}{a}sin(2\pi \varphi).
\end{equation}
It is the Euler equation of the variational principle with the action defined as
\begin{equation}
    S(\varphi)=Ta\int_{0}^{1}dt'\int_{-\infty}^{\infty}dx' \left[\frac{aM}{2T^2}\left(\frac{\partial \varphi(x',t')}{\partial t'}\right)^2-\frac{aG}{2}\left(\frac{\partial \varphi(x',t')}{\partial x'}\right)^2-\frac{\varepsilon}{2a}\left(1-cos\left(2\pi \varphi(x',t')\right)\right)\right].
\end{equation}
Feynman formulation of quantum field theory links classic systems with their quantum counterparts through usage of the classical action in the path integral defining the propagator \cite{Feyn}
\begin{equation}
    \langle\psi|e^{-\frac{i}{\hbar}TH}|\theta\rangle=\int_{C^{0,1}_{\theta,\psi}}D\phi(x',t')e^{-\frac{i}{\hbar}S(\phi)}
\end{equation}
with $H$ as quantum Hamiltonian 
%(if $|\theta\rangle$ is a wavefunction representing a state concentrated in point $\theta$) 
and $C^{0,1}_{\theta,\psi}$ as family of all continuous functions fulfilling boundary conditions: $\phi(x',0)=\theta(x'),\ \phi(x',1)=\psi(x')$.
Let us consider a diagonal element of the propagator for a static solution $\varphi$
\begin{equation}\label{prop}
    \langle\varphi|e^{-\frac{i}{\hbar}TH}|\varphi\rangle=\int_{C^{0,1}_{\varphi,\varphi}}D\phi(x',t')e^{-\frac{i}{\hbar}S(\phi)}.
\end{equation}
For further calculations, it is  convenient, to represent $\phi$ as
\begin{equation}
\phi=\varphi+\sum_i a_i\phi_i
\end{equation}
with $\phi_i$ as base vectors, that vanish at $t'=0$ and $t'=1$ to satisfy boundary conditions.
The Feynman integral can be evaluated by the use of stationary phase method \cite{Mas} (since any classical solution is a stationary point of the action functional)
\begin{equation}
    S(\phi)=S(\varphi)+\frac{1}{2}\sum_{j,k}a_j a_k \frac{\partial^2 S}{\partial a_j \partial a_k}(\varphi) + \ldots
\end{equation}
Second derivative of $S$ can be expressed as
\begin{equation}
    \frac{\partial^2 S}{\partial a_j \partial a_k}(\phi)= Ta\left( \phi_k ,\left(-\frac{aM}{T^2}\frac{\partial^2}{\partial t'^2}+aG\frac{\partial^2}{\partial x^2}- \frac{2\varepsilon\pi^2}{a}\cos\left(2\pi(\varphi+\sum_i a_i\phi_i)\right)\right)\phi_j\right)
\end{equation}
with scalar product defined as
\begin{equation}
    \left(\phi_k ,\phi_j\right)=\int_0^1 dt'\int dx' \phi_k\phi_j
\end{equation}
with limits on the spatial integral to be specified depending on particular problem. Inserting it back to the propagator \eqref{prop}   gives
\begin{equation}
    \langle\varphi|e^{-\frac{i}{\hbar}TH}|\varphi\rangle\simeq e^{-\frac{i}{\hbar}S(\varphi)}\int_{C^{0,1}_{0,0}}D\phi(x',t')e^{-\frac{i}{2\hbar}\sum_{j,k}a_j a_k Ta\left( \phi_k ,\left(-\frac{aM}{T^2}\frac{\partial^2}{\partial t'^2}+aG\frac{\partial^2}{\partial x^2}- \frac{2\varepsilon\pi^2}{a}\cos\left(2\pi\varphi\right)\right)\phi_j\right)}.
\end{equation}
It is important to note, that static solutions of the field equation are assumed to be eigenstates of the Hamiltonian (or adequate eigenstates are approximated by pure classical states)
\begin{equation}
    e^{-\frac{i}{\hbar}TE_q}\langle\varphi|\varphi\rangle\simeq e^{-\frac{i}{\hbar}S(\varphi)}\int_{C^{0,1}_{0,0}}D\phi(x',t')e^{-\frac{i}{2\hbar}\sum_{j,k}a_j a_k \left( \phi_k ,\left(-\frac{aM}{T}\frac{\partial^2}{\partial t'^2}+aGT\frac{\partial^2}{\partial x^2}- \frac{2\varepsilon\pi^2 T}{a}\cos\left(2\pi\varphi\right)\right)\phi_j\right)},
\end{equation}
where $E_q$ is the quantum energy of the state $\varphi$. Since $-\frac{a^2M}{T}\frac{\partial^2}{\partial t'^2}+a^2GT \frac{\partial^2}{\partial x^2}- 2\varepsilon\pi^2 T\cos\left(2\pi\varphi\right)$ is a sum of one-dimensional second-order linear Hermitian operators, it's eigenfunctions form a complete set of orthogonal functions. In this situation one can choose the set $\{\phi_i\}$ to be composed of said eigenfunctions
\begin{equation}
    e^{-\frac{i}{\hbar}TE_q}\langle\varphi|\varphi\rangle\simeq e^{-\frac{i}{\hbar}S(\varphi)}\int_{C^{0,1}_{0,0}}D\phi(x',t')e^{-\frac{i}{2\hbar}\sum_{j}a_j^2 \left( \phi_j ,\left(-\frac{aM}{T}\frac{\partial^2}{\partial t'^2}+aGT\frac{\partial^2}{\partial x^2}- \frac{2\varepsilon\pi^2 T}{a}\cos\left(2\pi\varphi\right)\right)\phi_j\right)}.
\end{equation}
Also $D\phi(x',t')$ can be rewritten as \cite{Byt}
\begin{equation}
    D\phi(x',t')=\prod_k da_k,
\end{equation}

\begin{equation}
    e^{-\frac{i}{\hbar}TE_q}\langle\varphi|\varphi\rangle\simeq e^{-\frac{i}{\hbar}S(\varphi)}\int_{-\infty}^{\infty}\prod_k da_k e^{-\frac{i}{2\hbar}\sum_{j}a_j^2 \left( \phi_j ,\left(-\frac{aM}{T}\frac{\partial^2}{\partial t'^2}+aGT\frac{\partial^2}{\partial x^2}- \frac{2\varepsilon\pi^2 T}{a}\cos\left(2\pi\varphi\right)\right)\phi_j\right)}.
\end{equation}
From orthogonality one obtains
\begin{equation}
    e^{-\frac{i}{\hbar}TE_q}\langle\varphi|\varphi\rangle\simeq e^{-\frac{i}{\hbar}S(\varphi)}\int_{-\infty}^{\infty}\prod_k da_k e^{-\frac{i}{2\hbar}a_k^2 \left( \phi_k ,\left(-\frac{aM}{T}\frac{\partial^2}{\partial t'^2}+aGT\frac{\partial^2}{\partial x^2}- \frac{2\varepsilon\pi^2 T}{a}\cos\left(2\pi\varphi\right)\right)\phi_k\right)}.
\end{equation}
If $\lambda_i$ will be considered as eigenvalues of $-\frac{a^2M}{T}\frac{\partial^2}{\partial t'^2}+a^2GT\frac{\partial^2}{\partial x'^2}- 2\varepsilon\pi^2 T\cos\left(2\pi\varphi\right)$, then
\begin{equation}
    e^{-\frac{i}{\hbar}TE_q}\langle\varphi|\varphi\rangle\simeq e^{-\frac{i}{\hbar}S(\varphi)}\int_{-\infty}^{\infty}\prod_k da_k e^{-\frac{i}{2\hbar}a_k^2 \lambda_k\left(\phi_k ,\phi_k\right)}.
\end{equation}
Substitution:
\begin{equation}
    a'_k=a_k \sqrt{\frac{\lambda_k\left(\phi_k ,\phi_k\right)}{2\hbar}},
\end{equation}
\begin{equation}
    da_k=da'_k \sqrt{\frac{2\hbar}{\lambda_k\left(\phi_k ,\phi_k\right)}},
\end{equation}

\begin{equation}
    e^{-\frac{i}{\hbar}TE_q}\langle\varphi|\varphi\rangle\simeq e^{-\frac{i}{\hbar}S(\varphi)}\prod_k\sqrt{\frac{2\hbar}{\lambda_k\left(\phi_k ,\phi_k\right)}} \int_{-\infty}^{\infty} da'_k e^{-ia'^2_k}.
\end{equation}
The remaining integral gives:
\begin{equation}
    \int_{-\infty}^{\infty} da'_k e^{-ia'^2_k}=\sqrt{\frac{\pi}{i}},
\end{equation}

\begin{equation}
    e^{-\frac{i}{\hbar}TE_q}\langle\varphi|\varphi\rangle\simeq e^{-\frac{i}{\hbar}S(\varphi)}\prod_k\sqrt{\frac{2\pi\hbar}{i\lambda_k\left(\phi_k ,\phi_k\right)}}.
\end{equation}
As for the norm of $\phi_k$
\begin{equation}
    \left(\phi_k ,\phi_k\right)=\int_0^1 dt'\int dx' \phi_k^2,
\end{equation}
it will be considered unitless with norm of $1$, which is ensured by previous change of variables (\ref{zamiana}). Thus:
\begin{equation}
    e^{-\frac{i}{\hbar}TE_q}\langle\varphi|\varphi\rangle\simeq e^{-\frac{i}{\hbar}S(\varphi)}\sqrt{\prod_k\frac{2\pi\hbar}{i\lambda_k}},
\end{equation}
\begin{equation}
    e^{-\frac{i}{\hbar}TE_q}\langle\varphi|\varphi\rangle\simeq e^{-\frac{i}{\hbar}S(\varphi)}\sqrt{\det\left[\frac{i}{2\pi\hbar}\left(-\frac{a^2M}{T}\frac{\partial^2}{\partial t'^2}+a^2GT\frac{\partial^2}{\partial x'^2}- 2\varepsilon\pi^2 T\cos\left(2\pi\varphi\right)\right)\right]}^{-1}.
\end{equation}
$E_q$ can be now expressed as
\begin{equation}
    E_q\simeq E(\varphi)+\frac{\hbar}{2iT}\ln \det\left[\frac{i}{2\pi\hbar}\left(-\frac{a^2M}{T}\frac{\partial^2}{\partial t'^2}+a^2GT\frac{\partial^2}{\partial x'^2}- 2\varepsilon\pi^2 T\cos\left(2\pi\varphi\right)\right)\right]-\frac{\hbar}{iT}\ln\langle\varphi|\varphi\rangle.
\end{equation}
For brevity
\begin{equation}\label{D}
    D=\frac{i}{2\pi\hbar} \left(-\frac{a^2M}{T}\frac{\partial^2}{\partial t'^2}+a^2GT\frac{\partial^2}{\partial x'^2}- 2\varepsilon\pi^2 T\cos\left(2\pi\varphi(x')\right)\right).
\end{equation}
Problem: differential operators have energy spectrum up to infinity, thus their determinants will be infinite as well. First step in solving the problem, is to properly set zero for the energy by subtracting an analogous term for vacuum \cite{ZL}
\begin{equation}\label{norm}
    E_q\simeq E(\varphi)+\frac{\hbar}{2iT}\ln \det\left[D\right]-\frac{\hbar}{2iT}\ln \det\left[D_0\right]-\frac{\hbar}{iT}\ln\langle\varphi|\varphi\rangle
\end{equation}
with
\begin{equation}
    D_0=\frac{i}{2\pi\hbar} \left(-\frac{a^2M}{T}\frac{\partial^2}{\partial t'^2}+a^2GT\frac{\partial^2}{\partial x'^2}\right).
\end{equation}
Second step is to rewrite the determinants in a form that allows subtraction. For this purpose, let us define generalised zeta-function as
\begin{equation}
    \zeta_{D}(s)=\sum_{k: \lambda_k >0}\lambda_k^{-s},
\end{equation}
where $\lambda_k$ are eigenvalues of a given operator $D$. One should note, that summation is done over eigenfunctions, also the summation itself should be treated in the most general sense. It is evident, that \cite{ZL}
\begin{equation}
    \ln\det[D]=-\frac{d \zeta}{ds}(0).
\end{equation}
To obtain $\zeta$ functions, when spectra of involved operators are unknown, one should consider an equation
\begin{equation}
    \left(\frac{\partial}{\partial y} + D\right)g_{D}\left(y,x',x'_{0}\right)=\delta\left(y \right)\delta\left(x'-x'_{0}\right).
\end{equation}
Green function for this equation can be written as (for $y>0$)
\begin{equation}
    g_{D}\left(y,x',x'_{0}\right)=\sum_{k}e^{-\lambda_{k} y}\psi_k\left(x'\right)\psi^{*}_k\left(x'_{0}\right)\Theta\left(y \right),
\end{equation}
where $\psi_k$ are orthonormal eigenfunctions of operator $D$.

Only positive eigenvalues of the operator $D$ are important for further calculations, so one can integrate over all spatial variables (the resulting function is defined as $\gamma$ function) and subtract $e^{0t}$ (0 is an eigenvalue for any linear operator):
$$\gamma_{D}\left(y\right)=-1+\int g_{D}\left(y,x,x\right)dx,$$
\begin{equation}
\gamma_{D}\left(y\right)=\sum_{k:\lambda_k >0}e^{-\lambda_{k} y}.
\end{equation}
By performing Mellin transform
$$\zeta_{D} (s)=\frac{1}{\Gamma(s)} \int_{0}^{+\infty}y^{s-1}\gamma_{D}(y)dy,$$
one obtains generalized Riemann zeta-function
\begin{equation}\label{refzeta}
    \zeta_{D} (s)=\sum_{k:\lambda_k >0} \lambda^{-s}_{k}.
\end{equation}

It is important to note, that renormalization done in (\ref{norm}) is not complete. There as yet arises an ultraviolet divergent term \cite{KF}, which can be countered by multiplication of the zeta-function by an appropriate term \cite{ZL}
\begin{equation}
	\zeta_r (s)=r^{2s}\zeta(s),
\end{equation}
which, from a point of view of the definition (\ref{refzeta}), may be interpreted as the spectrum rescaling $\lambda_k\rightarrow r^{-2}\lambda_k$.
Thus
\begin{equation}\label{ren}
    E_q\simeq E(\varphi)-\frac{\hbar}{2iT}\left(\zeta'(0)+2\ln(r)\zeta(0)\right)-\frac{\hbar}{iT}\ln\langle\varphi|\varphi\rangle.
\end{equation}

For an operator that can be written as a sum of operators dependant on different variables $D=D_1+D_2$, $\gamma$ function is a product of $\gamma$ functions of the summed operators \cite{ZL}
\begin{equation}
    \gamma_D=\gamma_{D_1}\gamma_{D_2},
\end{equation}
what greatly simplifies calculations.
In this work:
\begin{equation}
    D_1=\frac{i}{2\pi\hbar}\left(a^2GT\frac{\partial^2}{\partial x'^2}- 2\varepsilon\pi^2 T\cos\left(2\pi\varphi(x')\right)\right),
\end{equation}
\begin{equation}
    D_2=\frac{i}{2\pi\hbar}\left(-\frac{a^2M}{T}\frac{\partial^2}{\partial t'^2}\right).
\end{equation}

\section{Corrections for a Sin-Gordon kink}\label{Cor}
\indent We can now proceed to calculating energy corrections. For brevity let's define
\begin{equation}
	U(x')=\frac{2\pi^2\varepsilon}{a^2G}\cos\left(2\pi\varphi(x')\right)
\end{equation}
\begin{equation}
	U(x')=\frac{2\pi^2\varepsilon}{a^2G}\left(2k^2-1-2k^2\ cn^2\left(\sqrt{\frac{2\varepsilon}{a^2G}}\pi x';k\right)\right)
\end{equation}
as well as:
\begin{equation}
    A=\frac{ia^2GT}{2\pi\hbar},
\end{equation}
\begin{equation}
    B=\frac{ia^2M}{2\pi\hbar T},
\end{equation}
\begin{equation}
	m=\sqrt{\frac{2\varepsilon}{a^2G}}\pi .
\end{equation}
Having the potential $U$, one can find the Green function for the operator $D_1$:
\begin{equation}
\left(\frac{\partial}{\partial y} + A\frac{\partial^2}{\partial x^2} -AU(x)\right)\left(g_{D_1}\left(y,x,x_{0}\right)\right)=\delta(y)\delta\left(x-x_{0}\right).
\end{equation}
After rescaling $y_A=Ay$ we have
\begin{equation}
\left(\frac{\partial}{\partial y_A} + \frac{\partial^2}{\partial x^2} -U(x)\right)\left(g_{D_1}\left(\frac{y_A}{A},x,x_{0}\right)\right)=\delta(y_A)\delta\left(x-x_{0}\right).
\end{equation}
To obtain the Green function, one uses Laplace transform
\begin{equation}\label{a}
    \left(p + \frac{\partial^2}{\partial x^2} -U(x)\right)\hat{g}_{D_1}\left(p,x,x_{0}\right)=\delta\left(x-x_{0}\right),
\end{equation}
where
\begin{equation}
    g_{D_1}\left(\frac{y_A}{A},x,x_0\right) = \frac{1}{2\pi i}\int_l\hat{g}_{D_1}(p,x,x_0)e^{py_A}dp.
\end{equation}

For further calculations one needs only values of $\hat{g}_{D_1}$ for $x=x_0$. In such a situation $\hat{g}_{D_1}(p,x,x)=G(p,x)$, where $G(p,x)$ solves  equation, similar to Hermit one \cite{C.H},\cite{ZL}
\begin{equation}\label{Hermit}
    2GG'' - (G')^2 - 4(U(x)-p)G^2+1=0.
\end{equation}
To solve equation (\ref{Hermit}), one substitutes
\begin{equation}
    z=cn^2(mx;k),
\end{equation}
which gives:
\begin{equation}\label{f}
	\begin{array}{c}
     2G(G''_{zz} 4m^2(-k^2z^3+(2k^2-1)z^2+(1-k^2)z)+G'_z 2m^2(-3k^2z^2+(4k^2-2)z+1-k^2)) -\\ 4m^2(-k^2z^3+(2k^2-1)z^2+(1-k^2)z)\left(G'_z \right)^2 - 4(u(z)-p)G^2+1=0,
	\end{array}
\end{equation}
One postulates solution in a form
\begin{equation}
    G(p,z)=\frac{P(p,z)}{2\sqrt{Q(p)}},
\end{equation}
where P i Q are polynomials:
\begin{equation}
    P=p+az+b,
\end{equation}
\begin{equation}
    Q=q_3 p^3+q_2p^2+q_1p+q_0.
\end{equation}
After separating the equation with respect to monomials of p, one obtains:
\begin{equation}
    P(p,z)=p-m^2k^2z,
\end{equation}
\begin{equation}
    Q(p)=-p^3+m^2(2k^2-1)p^2+m^4k^2(1-k^2)p=-p(p-m^2k^2)(p-m^2(k^2-1)).
\end{equation}
For the kink case ($k=1$) function $G(p,x)$ takes the form
\begin{equation}
    G_{D_1}(p,x) = \frac{1}{2\sqrt{m^2-p}} - \frac{m^2sech^2(mx)}{2p\sqrt{m^2-p}}.
\end{equation}
Solution for a constant potential (vacuum) can be obtained from (\ref{f}) with constant $u(z)$
\begin{equation}
    G_{D_{0}}(p,x) = \frac{1}{2\sqrt{m^2-p}}.
\end{equation}
At this point one can perform subtraction from (\ref{norm})
\begin{equation}\label{G1}
    G_{D_{1}}-G_{D_{0}}=G_1(p,x) = -\frac{m^2sech^2(mx)}{2p\sqrt{m^2-p}}.
\end{equation}
Integrating with respect to $x$ one obtains
\begin{equation}
    \hat{\gamma}_1(p) = -\frac{m}{p\sqrt{m^2-p}},
\end{equation}
\begin{equation}
    \hat{\gamma}_1(p) = \frac{im}{p\sqrt{p-m^2}}.
\end{equation}
Inverse Laplace transform of the above function can be found in mathematical handbooks:
\begin{equation}
    \gamma_1\left(\frac{y_A}{A}\right) = Erf(im\sqrt{y_A}) =\frac{2}{\sqrt{\pi}}\int_0^{im\sqrt{y_A}}e^{-\chi^2}d\chi=
   \frac{2im\sqrt{y_A}}{\sqrt{\pi}}\int_0^1\exp[m^2y_A\tau^2]d\tau ,
\end{equation}
\begin{equation}
    \gamma_1(y) = \frac{2im\sqrt{Ay}}{\sqrt{\pi}}\int_0^1\exp[m^2Ay\tau^2]d\tau .
\end{equation}
Now one has to incorporate $\gamma_{D_2}$ calculated below:
$$\gamma_{D_2} (y)=\frac{1}{2\pi}\int_{\mathbb{R}}e^{-\textbf{k}^2 By}\textbf{dk},$$
$$\gamma_{D_2} (y)=\frac{1}{2\pi}\sqrt{\frac{\pi}{By}},$$
$$\gamma (y)=\gamma_1(y)\gamma_{D_2}(y),$$
\begin{equation}
\gamma (y)=\gamma_{1}(y)\gamma_{D_2}(y)=\frac{im\sqrt{A}}{\pi\sqrt{B}}\int_0^1\exp[m^2Ay\tau^2]d\tau.
\end{equation}
\begin{equation}
\gamma (y)=\gamma_{1}(y)\gamma_{D_2}(y)=\frac{m\sqrt{-A}}{\pi\sqrt{B}}\int_0^1\exp[-m^2(-A)y\tau^2]d\tau.
\end{equation}
After applying Mellin transform:
\begin{equation}\label{zetaA}
    \zeta(s)=\frac{m}{\pi\Gamma(s)}\sqrt{\frac{-A}{B}}\int_{0}^{+\infty} dy\ y^{s-1}\int_0^1 e^{-m^2(-A)y\tau^2}d\tau,
\end{equation}
$$\zeta(s)=\frac{m^{-2s+1}(-A)^{-s}}{\pi}\sqrt{\frac{-A}{B}}\int_0^1 d\tau \ \tau^{-2s}.$$
Above integral converges if $Re(s)<\frac{1}{2}$. With such an assumption one obtains
\begin{equation}\label{zetaB}
    \zeta(s)=\frac{m^{-2s+1}(-A)^{-s}}{\pi}\sqrt{\frac{-A}{B}}\frac{1}{-2s+1}.
\end{equation}
Derivative of $\zeta$ function:
\begin{equation}
    \zeta'(0)=\frac{m}{\pi}\sqrt{\frac{-A}{B}}\left(2-2\ln(m)-\ln(-A)\right).
\end{equation}
Inserting into (\ref{ren}) gives
\begin{equation}
        E_q\simeq a^2G\frac{16m^3}{3g}-\frac{\hbar}{2iT}\frac{m}{\pi}\sqrt{\frac{-A}{B}}\left(2-2\ln(m)-\ln(-A)+2\ln(r)\right) -\frac{\hbar}{iT}\ln\langle\varphi|\varphi\rangle.
\end{equation}
After substituting $m$, $g$, $A$ and $B$
\begin{equation}
    E_q\simeq \frac{1}{\pi}\sqrt{8\varepsilon a^2 G}-\frac{\hbar}{2}\sqrt{\frac{2\varepsilon}{a^2 M}}\left(2-2\ln\left(\sqrt{\frac{2\varepsilon}{a^2 G}\pi}\right)-\frac{3\pi}{2}i-\ln\left(\frac{a^2 GT}{2\pi \hbar}\right)+2\ln(r)\right)-\frac{\hbar}{iT}\ln\langle\varphi|\varphi\rangle.
\end{equation}
\begin{equation}
    E_q\simeq \frac{1}{\pi}\sqrt{8\varepsilon a^2 G}-\frac{\hbar}{2}\sqrt{\frac{2\varepsilon}{a^2 M}}\left(2-\ln\left(\frac{\varepsilon T}{\hbar}\right)-\frac{3\pi}{2}i+2\ln(r)\right)-\frac{\hbar}{iT}\ln\langle\varphi|\varphi\rangle.
\end{equation}
The choice of the scaling parameter $r$ allows to eliminate a dependence of the result on T (the simplest choice is $r^2=\frac{\varepsilon T}{\hbar}$).
\begin{equation}
    E_q\simeq \frac{1}{\pi}\sqrt{8\varepsilon a^2 G}+\hbar\sqrt{\frac{2\varepsilon}{a^2 M}}.
\end{equation}
This result for the correction coincides with one of \cite{KF}.
%The imaginary part gives $\langle\varphi|\varphi\rangle$. 
%It should be noted, that parameter $r$ leaves a degree of freedom allowing for normalization of %$\langle\varphi|\varphi\rangle$ as probability density. This direction won't be explored in this work, %since it focuses purely on corrections to energy.
\section{Dislocations}
\subsection{General scheme}\label{DGS}
\hspace{0.25in}An outline for modeling edge dislocations through Frenkel-Kontorova model consists of two steps \cite{Braun}. Firstly one simulates cross-section of the dislocation (in direction of Burgers vector) as a Frenkel-Kontorova kink. Parameters for the equation
\begin{equation}\label{FK2}
    M\frac{\partial^2 x_i}{\partial t^2}=G\left(x_{i+1}-2x_i+x_{i-1}\right)-\frac{\varepsilon \pi}{a}\sin\left(\frac{2\pi x_i}{a}\right)
\end{equation}
are obtained as follows: $M$ is the mass of atoms of which the particular crystal lattice is built, $a$ is the lattice constant, $G$ is proportional to the bulk modulus $K$ of a given material ($G=Ka$) and $\varepsilon$ is proportional to shear modulus $M_s$ ($\varepsilon=\frac{a^3 M_s}{2\pi^2}$).

At this point one can calculate energy of the dislocation and determine the Peierls-Nabarro potential \cite{Nabarro}, which is accountable for interactions between moving kinks and crystal lattice. It is important, not to approximate Frenkel-Kontorova equation with Sin-Gordon equation at this point, or else Peierls-Nabarro potential will vanish. Approximate solutions of Frenkel-Kontorova equation are often constructed on basis of Sin-Gordon soliton though.

    In the second step one describes the dislocation line as a row of Frenkel-Kontorova kinks that interact harmonically with their nearest neighbors and are subject to Peierls-Nabarro potential (or it's first term in Fourier expansion). This means, that the dislocation line is described by another Frenkel-Kontorova equation, which can be well substituted by Sin-Gordon equation. It is important to note, that the harmonic interaction coefficient as well as mass is not the same as in (\ref{FK2}).

    Movement of a dislocation is simulated through creation and propagation of second order kink-antikink pairs as a response to applied stress. Often the above described framework is enriched by taking into account additional terms of the potential or anharmonic interactions. It might also be important to consider thermal oscillations of the lattice in modeling macroscopic properties of crystals.
\subsection{Calculations}
\hspace{0.25in} Peierls-Nabarro potential's amplitude ($\varepsilon_2$) and harmonic interaction coefficient ($G_2$) for dislocation line modeling were calculated numerically. Firstly approximate static Frenkel-Kontorova kinks were obtained by substituting Sin-Gordon kink as an initial condition to a critically damped Frenkel-Kontorova equation (already in finite difference form)
\begin{equation}
f(x_i,t_j)=f(x_i,t_{j-1})+(f(x_{i+1},t_{j-1})-2f(x_i,t_{j-1})+f(x_{i-1},t_{j-1}))dt+\frac{\varepsilon \pi}{a^2 G}\sin(2\pi f(x_i,t_{j-1}))dt.
\end{equation}

By changing initial position of the kink, one gets two different stable Frenkel-Kontorowa kinks, which correspond to minimal and maximal kink energy. Difference of their energy gives $\varepsilon_2$. $G_2$ is estimated on assumption, that energy of any two atoms interacting is given by
\begin{equation}
E(r)=\frac{C_1}{r^2}-\frac{C_2}{r},
\end{equation}
where $C_1$ and $C_2$ are fitted so that $E(a)$ is the minimum and $E''(a)=G$
\begin{equation}
\left\{\begin{array}{cc}
C_1=\frac{1}{2}a^4 G \\
C_2=a^3 G
\end{array}\right. .
\end{equation}

Interaction energy for a given displacement between kinks ($\Delta X$) is then calculated by adding up interaction energy of atom pairs from two neighbouring kinks. To avoid major rounding up inaccuracy and set in the zero for energy, one subtracts an analogous term for $\Delta X=0$ on every iteration. By varying $\Delta X$, one can obtain $G_2$ by fitting $\frac{G_2}{2}\Delta X^2$ as the energy. It is important, to fit this function separately for both Frenkel-Kontorova kinks, which means that $\Delta X=na$.
Effective mass of the kink is obtained thanks to the fact, that Sin-Gordon equation is Lorentz invariant (with sound velocity $c=\sqrt{\frac{a^2 G}{M}}$ used instead of speed of light in vacuum)
\begin{equation}
    M_2=\frac{E_0}{c^2}=\frac{6}{\pi}M\sqrt{\frac{2\varepsilon}{a^2 G}},
\end{equation}
where $E_0$ is the energy of a static kink.
\subsection{Application}
\hspace{0.25in} Aim for this subsection is to estimate energy corrections to a kink-antikink pair approximating it with a sum of kink and antikink (which is acceptable when kink and antikink are far enough from each other). This in turn will have effect on calculated plasticity of a given material \cite{Num}. As pointed in section (\ref{DGS}) all important coefficients can be obtained from bulk modulus, shear modulus and lattice parameters for a given crystal. It is to be noted, that one can only estimate energy correctly for pure elements monocrystals. Further calculations in this subsection are based on parameters of polycrystal metals (due to lack of more relevant sources), thus are only estimation of order of quantum corrections for kink energy in dislocations. As can be seen in Table \ref{tab2:} ($E_d$ as classical energy and $\Delta E_d$ as corrections), corrections are three orders of magnitude lower then the classical energy. Therefore they should be considered if desired precision is of three significant figures or more.
\section{Crowdions}
\hspace{0.25in} The most basic model of a crowdion \cite{Braun} is similar to the first step of dislocation modelling. Essentially the same Frenkel-Kontorova equation is used with the exception, that the potential now describes interaction with all atoms surrounding the modelled chain of atoms. For the purpose of this section the effect of external stress on crowdion energy will not be included as well as deformation of the lattice around the crowdion (on contrary to \cite{CR} or \cite{CR2})
\begin{equation}\label{crowdion}
    M\frac{\partial^2 x_i}{\partial t^2}=G\left(x_{i+1}-2x_i+x_{i-1}\right)-\frac{\varepsilon \pi}{a}\sin(\frac{2\pi x_i}{a}),
\end{equation}
where $M$ is the mass of atoms of which the particular crystal lattice is built, $a$ is the, $G$ is proportional to the bulk modulus $K$ of a given material ($G=Ka$) and $\varepsilon$ is proportional to shear modulus $M_s$ ($\varepsilon=\frac{2a^3 M_s}{\pi^2}$ - only interaction with nearest neighbors is taken into account). The above equation is approximated by Sin-Gordon equation as described in section (\ref{FKSG})
\begin{equation}
\frac{M}{a}\frac{\partial^2 u}{\partial t^2}-a G\frac{\partial^2 u}{\partial x^2}+\frac{\varepsilon \pi}{a^2}\sin\left(\frac{2\pi u}{a}\right)=0.
\end{equation}
Crowdion classical energy and quantum corrections are calculated as shown in sections (\ref{FKSG}) and (\ref{Cor}).

\begin{table}[h]
\scriptsize
\centering
\begin{tabular}{|c|c|c|c|c|c|c|c|c|}\hline
& M [kg $\cdot 10^{-26}$] \cite{Wol} &$a$ [nm] \cite{Por}& $M_s$ [GPa] \cite{Wol} & $K$ [GPa] \cite{Wol} & $E_d$ [meV] & $\Delta E_d$ [meV] &$E_c$ [eV] & $\Delta E_c$ [eV] \\ \hline
Ag & $17,9119$ & $0,40776$ & $30$ & $100$ & $1,9229$ & $0,0113$ & $9,3934$ & $0,0034$\\ \hline
Al & $4,48345$ & $0,40494$ & $26$ & $76$ & $2,7443$ & $0,0336$ & $7,4665$ & $0,0064$\\ \hline
Au & $32,7071$ & $0,40702$ & $27$ & $180$ & $0,06719$ & $0,00033$ & $11,8909$ & $0,0024$\\ \hline
Cu & $10,552$ & $0,3610$ & $48$ & $140$ & $3,1938$ & $0,0283$ & $8,7556$ & $0,0053$\\ \hline
Fe & $9,2733$ & $0,28665$ & $82$ & $170$ & $7,4299$ & $0,0996$ & $7,0290$ & $0,0066$\\ \hline
Mg & $4,03594$ & $0,5196$ & $17$ & $45$ & $5,5603$ & $0,0449$ & $9,8149$ & $0,0061$\\ \hline
Ni & $9,7463$ & $0,429$ & $76$ & $180$ & $14,4136$ & $0,0786$ & $23,3596$ & $0,0076$\\ \hline
\end{tabular}
\caption{Quantum corrections to dislocation kink and crowdion energy for exemplary metals}
\label{tab2:}
\end{table}

    Corrections (see Table \ref{tab2:}) are four orders of magnitude lower then the classical energy, yet they can be notable in precise measurements.

\section{Conclusion}
An important element of the approach which we demonstrate in the Sec. 4 relates to the renormalization problem. Account of the "vacuum" contribution with the constant potential term gives the finite result for the energy correction (compare with  e.g. \cite{Mach}).

\end{document}